\documentclass[prl,aps,twocolumn,superscriptaddress,nofootinbib]{revtex4-2}
\usepackage{dcolumn}
\usepackage{bm}
\usepackage{graphicx}
\usepackage{color}
\usepackage{subfigure}
\usepackage{hyperref}
\usepackage{latexsym}
\usepackage{amsthm}
\usepackage{amssymb}
\usepackage{array}
\usepackage{braket}
\usepackage{subfigure}
\usepackage{amsmath,amssymb}
\usepackage{bbm}
\usepackage{booktabs}
\usepackage{subfigure}
\usepackage{mathrsfs}
\usepackage{warpcol}
\usepackage{amsthm}

\usepackage{subfigure}

\usepackage{lipsum}
\usepackage{graphics}
\usepackage[dvipsnames]{xcolor}
\DeclareGraphicsExtensions{.jpg,.pdf, .mps, .png, .eps, .ps, .EPS,.gif}

\begin{document}

\def\be{\begin{equation}}
\def\ee{\end{equation}}
\def\bc{\begin{center}}
\def\ec{\end{center}}
\def\bea{\begin{eqnarray}}
\def\eea{\end{eqnarray}}
\newcommand{\avg}[1]{\langle{#1}\rangle}
\newcommand{\Avg}[1]{\left\langle{#1}\right\rangle}
\def\ie{\textit{i.e.}}
\def\etal{\textit{et. al.}}
\def\m{\vec{m}}
\def\G{\mathcal{G}}
\newcommand{\gin}[1]{{\bf\color{blue}#1}}
\newcommand{\kir}[1]{{\bf\color{PineGreen}#1}}
\newtheorem{theorem}{Theorem}
\newtheorem{corollary}{Corollary}
\newtheorem{lemma}{Lemma}
\newtheorem{conjecture}{Conjecture}
\newtheorem{proposition}{Proposition}
\newcommand{\tr}{\ensuremath{\tilde{\bm\rho}}}
\newcommand{\dsi}{\ensuremath{\Delta\bm\sigma_i}}

\newcommand{\Bruce}[1]{{\color{blue}#1}}

\title{Contrarians synchronize beyond the limit of pairwise interactions}

\author{K. Kovalenko$^{+,*,}$}
\affiliation{Moscow Institute of Physics and Technology, Dolgoprudny, Moscow Region, 141701, Russian Federation}

\author{X. Dai$^{+,}$}
\affiliation{Northwestern Polytechnical University, Xi'an 710072, P. R. China}

\author{K. Alfaro-Bittner$^{+,}$}
\affiliation{Departamento de F{\'i}sica, Universidad T{\'e}cnica Federico Santa Mar{\'i}a, Av. Espa{\~n}a 1680, Casilla 110V, Valpara{\'i}so, Chile}

\author{A. M. Raigorodskii}
\affiliation{Moscow Institute of Physics and Technology, Dolgoprudny, Moscow Region, 141701, Russian Federation}
\affiliation{Adyghe State University, ul. Pervomaiskaya, 208, Maykop, 385000, Russia}
\affiliation{Moscow State University, Leninskie Gory, 1, Moscow, 119991, Russia}
\affiliation{Buryat State University, ul. Ranzhurova, 5, Ulan-Ude, 670000, Russia}

\author{M. Perc}
\affiliation{Faculty of Natural Sciences and Mathematics, University of Maribor, Koro{\v s}ka cesta 160, 2000 Maribor, Slovenia}
\affiliation{Department of Medical Research, China Medical University Hospital, China Medical University, Taichung 404332, Taiwan}
\affiliation{Complexity Science Hub Vienna, Josefst{\"a}dterstra{\ss}e 39, 1080 Vienna, Austria}
\affiliation{Alma Mater Europaea, Slovenska ulica 17, 2000 Maribor, Slovenia}

\author{S. Boccaletti}
\affiliation{Moscow Institute of Physics and Technology, Dolgoprudny, Moscow Region, 141701, Russian Federation}
\affiliation{Universidad Rey Juan Carlos, Calle Tulip\'an s/n, 28933 M\'ostoles, Madrid, Spain}
\affiliation{CNR - Institute of Complex Systems, Via Madonna del Piano 10, I-50019 Sesto Fiorentino, Italy}

\begin{abstract}
	We give evidence that a population of pure contrarians globally coupled $D$-dimensional Kuramoto oscillators reaches a collective synchronous state when the interplay between the units goes beyond the limit of pairwise interactions.
	Namely, we will show that the presence of higher order interactions may induce the appearance of a coherent state even when the oscillators are coupled negatively to the mean field.
	An exact solution for the description of the microscopic dynamics for forward and backward transitions is provided, which entails imperfect symmetry breaking of the population into a frequency-locked state featuring two clusters of different instantaneous phases.
	Our results contribute to a better understanding of the powerful potential of group interactions entailing multi-dimensional choices and novel dynamical states in many circumstances, such as in social systems.
\end{abstract}
\maketitle

$^{*}$ Corresponding Author: kkd15@mail.ru

$^{+}$ These Authors equally contributed to the Manuscript

When Kuramoto introduced its model of globally coupled phase oscillators~\cite{winfree1980geometry,kuramoto1984chemical}, he certainly could have hardly fathomed that it would have acquired such a broad applicability over the decades~\cite{strogatz2000kuramoto, acebron2005kuramoto, arenas2008synchronization, pikovsky2015dynamics, rodrigues2016kuramoto, boccaletti2016explosive, parastesh_pr21} to systems of physical interest, such as Josephson junctions~\cite{swift1992averaging, wiesenfeld1996synchronization}, laser arrays~\cite{braiman1995entrainment, kourtchatov1995theory, kozyreff2000global}, oscillator glass~\cite{daido1992quasientrainment, daido2000algebraic}, and charge density waves~\cite{gruner1988dynamics}.
The model became soon a paradigm for the study of synchronization, i.e. the emergent property through which a system forms a collective, coherent, rhythm~\cite{boccaletti2016explosive,yeung_time_1999,strogatz_kuramoto_2000} via the interaction of its oscillatory units, which has been the object of a wide range of researches in physics, biology, and engineering~\cite{boccaletti_synchronization_2002,pikovskij_synchronization_2003,boccaletti_synchronization_2018}.
In the Kuramoto model, a necessary condition for synchronization is to have a rigorously non negligible fraction (usually larger than a finite threshold) of oscillators which are acting as conformists, i.e. which are featuring {\it a strictly positive} coupling strength to the mean field. The emergence of order for repulsive strengths has indeed been observed so far only in specific, local, coupling arrangements \cite{ionita_kuramoto2013,Shadisadat_2017}.

In social science, various generalizations of the Kuramoto model were proposed for the study, for instance, of human crowd behavior during clapping~\cite{neda2000sound, neda2000physics} and crossing bridges~\cite{strogatz2005crowd, eckhardt2007modeling}, and of human decision making in general. The inclusion of conformists and contrarians as positively and negatively coupled individuals to the mean field captures well the essential dichotomy of many social interactions and their relation to whatever the prevailing opinion might be~\cite{hong2011kuramoto, hong2011conformists}. It was also argued that the continuous spectrum of opinions as points on a circle reflect the political reality better than the traditionally considered linear continuum from the left to the right wing~\cite{durlauf2004social}.
Applications to social interactions of the Kuramoto model have not received, however, as much attention as it would have deserved. This may be due to two fundamental limitations of the original model. In the first place, social interactions typically unfold in more than two dimensions, simply because the spectrum of behavioral choices often goes beyond a circle representation. It has recently been noted that the quest for moral behavior in an evolutionary setting, for example, may entail choosing between many different strategies, all of which with highly nonlinear consequence for the individual and the social network as a whole~\cite{capraro2020lying, capraro_jrsi21}. Such a decision space thus surely requires more than a circle to be completely mapped out. And secondly, social interactions inherently entail groups that are not accurately described by pairwise links in traditional oscillator networks~\cite{alvarez-rodriguez_nhb21}. Cooperation in large groups of unrelated individuals, for example, distinguishes us most from other mammals, and it is one of the central pillars of our evolutionary success~\cite{nowak_11, perc_pr17}. But classical social networks with only pairwise links simply do not provide a unique procedure for defining a group~\cite{battiston2020networks}.

\begin{figure}[t!]
	\centering
	\includegraphics[scale=0.8]{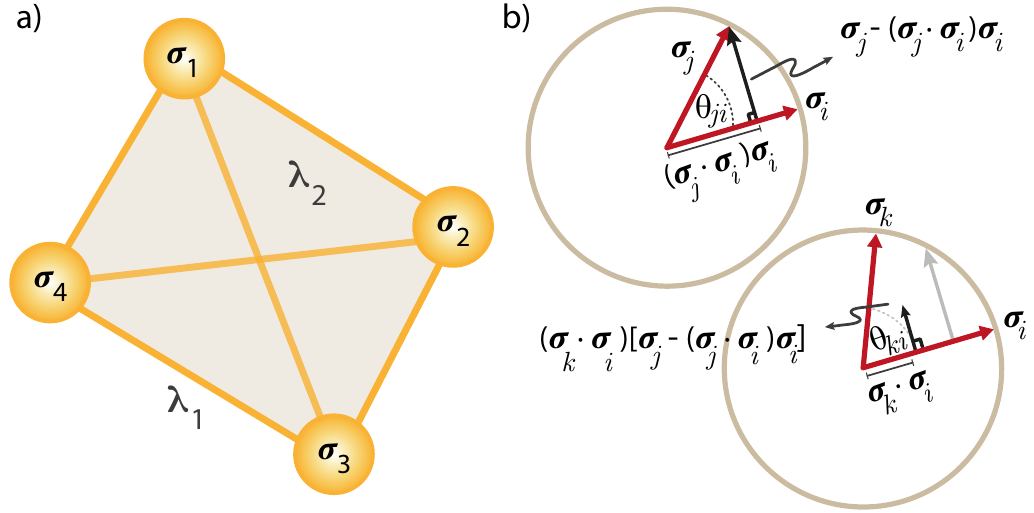}
	\caption{(a) The setting of Eq.~(\ref{eq:vectorial_long}): D-dimensional oscillators are represented by yellow circles, and they interact via pairwise and
		triadwise interactions, with coupling strengths respectively given by $\lambda_1$ and $\lambda_2$. (b) $D=2$. The various trigonometric relationships which
		allow one to transform of the classical Kuramoto model (Eq.~(\ref{eq:mixed_D=2})) into Eq.~(\ref{eq:vectorial_long}).}
	\label{fig:01}
\end{figure}

\begin{figure*}[th]
	\centering
	\includegraphics[scale=0.75]{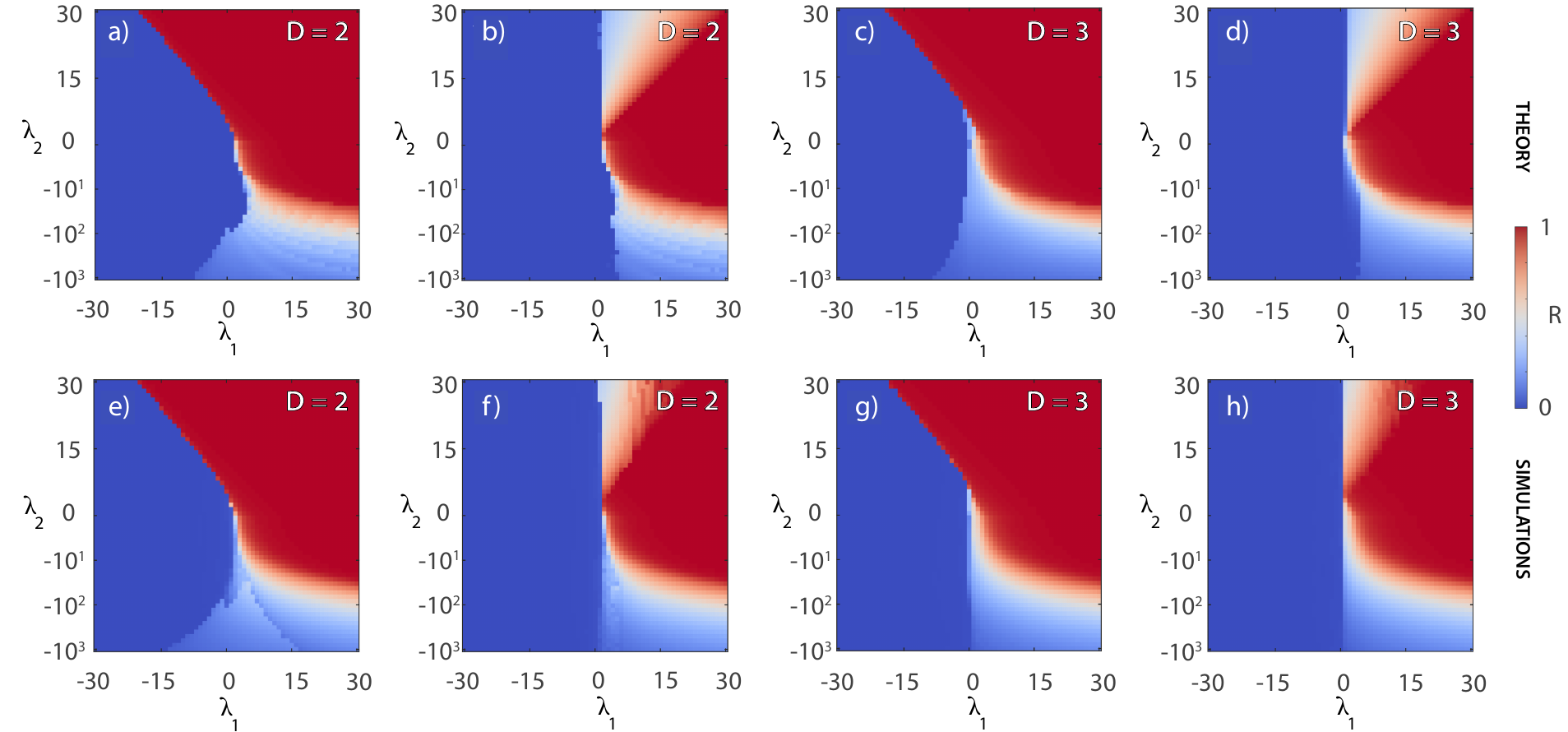}
	\caption{Theoretical predictions [panels (a-d), calculated from Eq.~(26) of the SI] and numerical simulations [panels (e-h), calculated from Eq.~(\ref{eq:vectorial_long})] for $D = 2$ and $D=3$ of the order parameter $R$ (see text for definitions) as a function of the coupling constants $\lambda_1$ and $\lambda_2$.
		Panels (a,c,e,g) refer to the backward transition; panels (b,d,f,h) refer to the forward transition. In panels (a-d), the values around the line $\lambda_1=-\lambda_2$ of the quadrant defined by $\lambda_1>0$ and $\lambda_2<0$ are interpolated (see the SI for a full discussion on the theoretical limitations occurring at $\lambda_1=-\lambda_2$ for the assessment of stability of the synchronous state).
		In each panel, the synchronized (incoherent) state is represented by the red (blue) color, and the values of $R$ are coded according to the vertical color bar
		reported at the rightmost of the Figure. Notice that, for a better representation of the results, a vertical logarithmic scale is adopted for $\mid \lambda_2 \mid$ in the semi plane $\lambda_2<0$.  See the main text for the details on the size and on the initialization of the system.}
	\label{fig:02}
\end{figure*}

Aiming to overcome these limitations, in this Letter we consider a $D$-dimensional Kuramoto model with one and two-simplex interactions, and which includes both conformists and contrarians as positively and negatively coupled individuals to the mean field to capture the fact that some individuals -- the conformists -- prefer to go with the main stream, while others -- the contrarians -- prefer to oppose it. We will show that these generalizations allow us to retain full analytical tractability of the model, whilst also yielding fundamentally novel behavior. Notably, in addition to a rich plethora of behavior previously associated with various Kuramoto models~\cite{moreno2004synchronization, gomez2011explosive, zhang2015explosive, tinsley2012chimera, bi2016coexistence, chandra2019continuous}, such as mono- and multi-stability, spontaneous symmetry breaking, and explosive (i.e. discontinuous) transitions to synchronization, we give evidence that synchronization may occur also in the complete absence of conformists, a result which is {\it inherently prohibited} when only pairwise interactions take place in the ensemble. Thus, even if everybody contests the prevailing attitude, consensus is possible due to higher-order structures in the population.

Let us then start by considering an ensemble of globally coupled $N$ oscillators satisfying
\bea
\dot{\theta}_i=\omega_i+\dfrac{\lambda_1}{N}\sum_{j=1}^{N}\sin(\theta_j-\theta_i)+ \nonumber\\
+\dfrac{\lambda_2}{2N^2}\sum_{j=1}^{N}\sum_{k=1}^{N}\sin(\theta_j+\theta_k-2\theta_i),
\label{eq:mixed_D=2}
\eea
where $\theta_i$ is the phase and $\omega_i$ the natural frequency of each oscillator $i$ ($i=1,2\ldots,N$), while $\lambda_1$ and $\lambda_2$ are two real parameters accounting for the coupling strengths of pairwise and triadwise interactions, respectively, as schematically shown in panel (a) of Fig. \ref{fig:01}.
Now, one can consider the unit vector $\bm{\sigma}_i= \left( \cos(\theta_i),\sin(\theta_i) \right)$, and the antisymmetric matrix
$\bm{W}_i=
\begin{pmatrix}
	0 & \omega_i \\
	-\omega_i & 0
\end{pmatrix}.$
With the help of a few trigonometric relationships [which are schematically reported in panel (b) of Fig. \ref{fig:01}], and after denoting by $\bm\rho$ the mean vector of all $\bm\sigma_i$ (i.e. $\bm\rho = \frac{1}{N}\sum_{i=1}^{N}\bm\sigma_i$), Eq.~\ref{eq:mixed_D=2} can be rewritten as

\bea
\dot{\bm\sigma_i}=\bm{W}_i\bm\sigma_i+\lambda_1[\bm\rho-(\bm\rho\cdot\bm\sigma_i)\bm\sigma_i]+\nonumber\\
\lambda_2 (\bm\rho\cdot\bm\sigma_i)[\bm\rho-(\bm\rho\cdot\bm\sigma_i)\bm\sigma_i].
\label{eq:vectorial_long}
\eea

In fact, there is no reason for limiting Eq.~(\ref{eq:vectorial_long})  to the dynamics of 2-dimensional vectors.
On the contrary, the same equation can be adopted to describe the evolution of $D$-dimensional vectors $\bm\sigma_i$ (of norm one) whose trajectories lie on a $(D-1)$-dimensional unit sphere ${\cal{S}}^{D-1}$.
In the special case of $\lambda_1=\lambda_2=0$ equal to zero (i.e., where there are no interactions among the oscillators), $\bm\sigma_i$ rotates independently along some trajectory on ${\cal{S}}^{D-1}$ dictated by a real anti-symmetric matrix $\bm{W}_i\in\mathbb{R}^{\text{D}\times D}$ which, in what follows, is independently drawn at random for each node $i$. Specifically, each upper triangular element of  $\bm{W}_i$ is sampled from a Gaussian distribution $\mathcal{N}(0,1)$, and the lower-triangular elements are accordingly fixed to make ${\bm W}_i$ anti-symmetric.

For $D=2$ and $D=3$, Eq.~(\ref{eq:vectorial_long}) admits a rigorous analytical treatment, which ultimately yields a self-consistent equation for the order parameter $R \equiv  \langle \mid \bm\rho \mid \rangle_T$ (with $\langle...\rangle_T$ indicating time average over a sufficiently long time span $T$, and $\mid ... \mid$ indicating the norm).
Such a self-consistent equation allows, on its turn, the computation of the stationary points (and of their stability) for all values of $\lambda_1$ and $\lambda_2$.
The interested reader can find the complete details of the treatment in the Supplementary Information (SI).
On the other hand, we performed large scale simulations of Eq.~(\ref{eq:vectorial_long}), at $D=2$ and $D=3$, in an ensemble of $N = 5,000$ oscillators, by the use of a fourth-order Runge-Kutta algorithm with integration time $h = 10^{-3}$. In our simulations, we monitored both the forward and the backward transition to synchronization.
In the forward (backward) transition, the system is initialized in the fully incoherent (fully coherent) state, which corresponds to $R(t=0)\sim 0$ ($R(t=0)=1$), and, after a suitable transient time, the asymptotic state of the order parameter $R$ is calculated. In practice, this is realized by selecting a random unitary vector $e_D \in \mathbb{R}^D$, and letting each oscillators $\bm\sigma_i$ to be initially equal to $e_D$ with probability $0 \leq \mu \leq 1$, while setting $\bm\sigma_i(t=0)=-e_D$ with probability $1-\mu$. Then, the case $\mu=1$ ($\mu=0.5$) gives the proper initial conditions for inspection of the backward (forward) transition to synchronization.

We start our discussion by reporting in Figure \ref{fig:02} the theoretical predictions [panels of the first row, obtained by solving Eq.~(26) of the SI] and the numerical simulations [panels of the second row, obtained from Eq.~(\ref{eq:vectorial_long})], at $D = 2$ (first two columns) and $D=3$ (second two columns), for $R$ in the parameter plane ($\lambda_1$, $\lambda_2$). Precisely, panels a,c,e,g refers to the backward transition, whereas panels b,d,f,h reports the results of the forward transition to synchronization.

Figure \ref{fig:02} is informative on many relevant dynamical scenarios supported by system (\ref{eq:vectorial_long}). Some of these scenarios confirm and actually extend findings already reported in the literature: for instance the fact that triadwise interactions are detrimental for the forward transition, as in the limit of $\lambda_1=0$ the forward transition disappears, and in general the threshold in $\lambda_1$ needed to produce the synchronous state from incoherence increases monotonically with $\lambda_2$.
(see panels b,d,f,h) \cite{Dai_Kovalenko,arenaskura}.
Some other scenarios, instead, point to novel and relevant features of the system, which certainly deserve more detailed investigations:
for instance the fact that, despite the simplicity of the model, the simple combination between pure 1 and 2-simplices make it possible for the system
to exhibit mono-stability, bi-stability and multi-stability regions, as well as the fact that in the region of negative $\lambda_1$ and positive $\lambda_2$ the system features an abrupt transition from the fully synchronized state to the unsynchronized one.
Furthermore, the theoretical predictions (contained in the SI) and the numerical results are in very good agreement, and therefore the developed theory allows to unveil the origin of bi-stability in the Kuramoto model, as reported also in various other settings and circumstances  \cite{Dai_Kovalenko,arenaskura}.

But the most remarkable evidence that Figure~\ref{fig:02} is communicating is the presence of synchronization features in the backward transition for negative values of the coupling strengths.  Precisely, for $D=2$ and $D=3$ the theory (panels a and c, respectively) predicts the emergence of synchronization even when $\lambda_1$ and $\lambda_2$ are both negative, a dynamical state which actually would be inherently prohibited if the interplay among the elements would be limited to pairwise interactions. In other words, group interactions lead to the emergence of synchronization, or in social terms to the emergence of agreement, even if all individuals in the network are contrarians, i.e., they are opposing the mainstream at all times~\cite{hong2011kuramoto}. In simulations, we have observed such state only for $D=2$ (i.e. for the classical Kuramoto model), whereas a detailed analysis of the $D=3$ case will be reported elsewhere.

For a better visualization of this latter macroscopic and generic effect, in the following we will fix $D=2$, and focus into the microscopic details of the observed synchronization of {\it pure contrarians}.
To that purpose, in Figure \ref{fig:04} we report four time snapshots [panels a,b,c,d)] of the vectors \{ $\bm\sigma_i \equiv \exp(i \theta_i)$ \} (with norms slightly adjusted around unity for a better visibility) in the plane ($\sigma_x = \cos\theta$, $\sigma_y=\sin\theta$), together with the oscillators' instantaneous phases  [panels a1,b1,c1,d1] and frequencies [panels a2,b2,c2,d2] as functions of their natural frequency $\omega_i$.
It is seen that the system sets actually in a {\it collective state}: starting from a fully incoherent dynamics ($\lambda_1=-10,\lambda_2=-10^2$), where the unit vectors are almost homogenously distributed around the unit circle (panel a), with unlocked phases (panel a1) and frequencies (panel a2), a progressive increase in $\lambda_1$ determines
a symmetry breaking scenario wherein two groups of oscillators form (panels b,b1 and b2, obtained at $\lambda_1= 0, \lambda_2=-10^2$), and grow in correspondence with decreasing $\lambda_2$, up to the final organization of the population of contrarians (panels c, c1, and c2, obtained for $\lambda_1=-10, \lambda_2= -10^3$). In this latter state, almost all oscillators display instantaneous phases within either one of the two phase clusters, whose centers of mass are located at positions $\tilde \phi(t)$ and $\pi-\tilde \phi (t)$ respectively (as it can be seen in panel c1) and rotate coherently with a fully locked frequency (see panel c2). As a consequence of the two clusters, the value of the order parameter remains small, and yet distinguishable from that corresponding to the fully incoherent dynamics.
Moreover, it is very interesting to notice that, while the presence of only pairwise interactions would determine the total destruction of such a state, when pairwise and triadwise interactions are simultaneously present, positive values of $\lambda_1$ may actually enhance the dichotomic nature of the state (see panels d, d1, and d2).
We refer the interested reader to the discussion of our SI, where we prove that for $D=2$ and $\lambda_2=0$ (i.e., for the classical case of the Kuramoto model with only pairwise interactions), if $\lambda_1$ is negative, the only solution of the self-consistent equation is $R=0$ (i.e., synchronization of contrarians is prevented in this case), while the presence of 3-body interactions introduces a new solution of the self-consistent equation (the observed new state) even for $\lambda_1<0$ and $\lambda_2<0$.

\begin{figure}[t]
	\centering
	\includegraphics[scale=0.7]{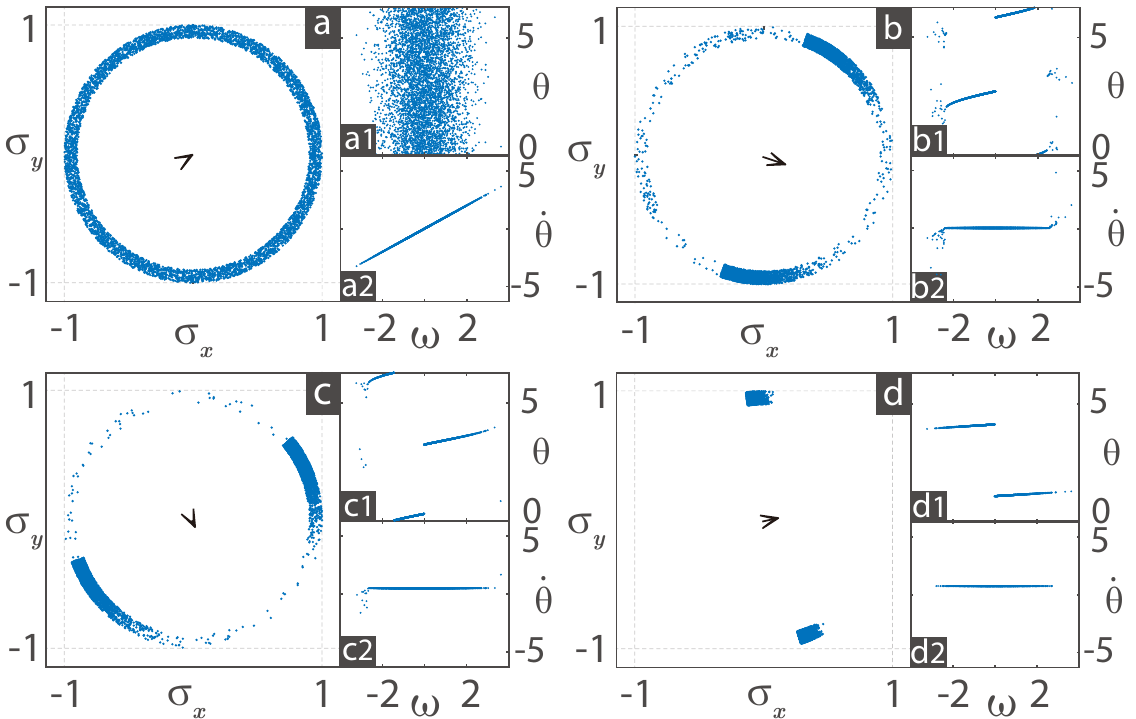}
	\caption{The microscopic details behind the synchronization of contrarians.
		Panels (a-d) report (with blue dots) a time snapshot of the set of vectors \{ $\bm\sigma_i \equiv \exp(i \theta_i)$ \} (for better visibility the norms of such vectors have been
		slightly adjusted around unity) in the plane ($\sigma_x = \cos\theta$, $\sigma_y=\sin\theta$). The black arrow stands for the instantaneous order parameter vector $\bm\rho \equiv \sum_{i=1}^{N} \exp(i \theta_i)$. (a) $\lambda_1=-10,\lambda_2=-10^2$; (b) $\lambda_1= 0, \lambda_2=-10^2$; (c) $\lambda_1=-10, \lambda_2= -10^3$; (d) $\lambda_1= 10, \lambda_2= -10^3$. In all cases $\mu=1$ for the initialization of the system. Panels a1,b1,c1,d1 (a2,b2,c2,d2) report the instantaneous phase $\theta_i$ (the instantaneous frequency $\dot \theta_i$) for each oscillator $i$ in the ensemble, as a function of the oscillator's natural frequency $\omega_i$.}
	\label{fig:04}
\end{figure}

Finally, we report that the emergence of such collective state does not necessarily require an all-to-all configuration, but, on the contrary, it can be observed also is sparser connectivity structures. To this purpose, we set $N=10^2$ and simulate Eq.~(\ref{eq:mixed_D=2}) for $\lambda_1=-10, \lambda_2=-1000$, starting from an all-to-all configuration, and initializing the phases with $\mu=1$. As a consequence, the system sets in the collective state reported in panel (a) of Fig.~\ref{fig:04C}. Then, every 50 time steps, a fraction $1-p$ of triangular interactions is randomly removed from the double sum of the second term of the right hand side of Eq.~(\ref{eq:mixed_D=2}).
Panel (d) of Fig. \ref{fig:04C} reports then the behavior of $R_1=\dfrac{1}{N}\|\sum_{j=1}^{N}e^{i\theta_j}\|$ (red curve with squares) and $R_2=\dfrac{1}{N}\|\sum_{j=1}^{N}e^{2i\theta_j}\|$ (blue curve with circles) vs. the fraction $p$ of remaining triangles. It is clearly seen that the new collective state is resilient up to the removal of about 20\% of triadwise interactions [see panel (b) of Fig.~\ref{fig:04C}], and eventually disappears completely when $p=1$, in full agreement with the rigorous result that a population of pure contrarians is strictly prevented from synchronizing in the limit of pairwise interactions.

\begin{figure}[t]
	\centering
	\includegraphics[width=\linewidth]{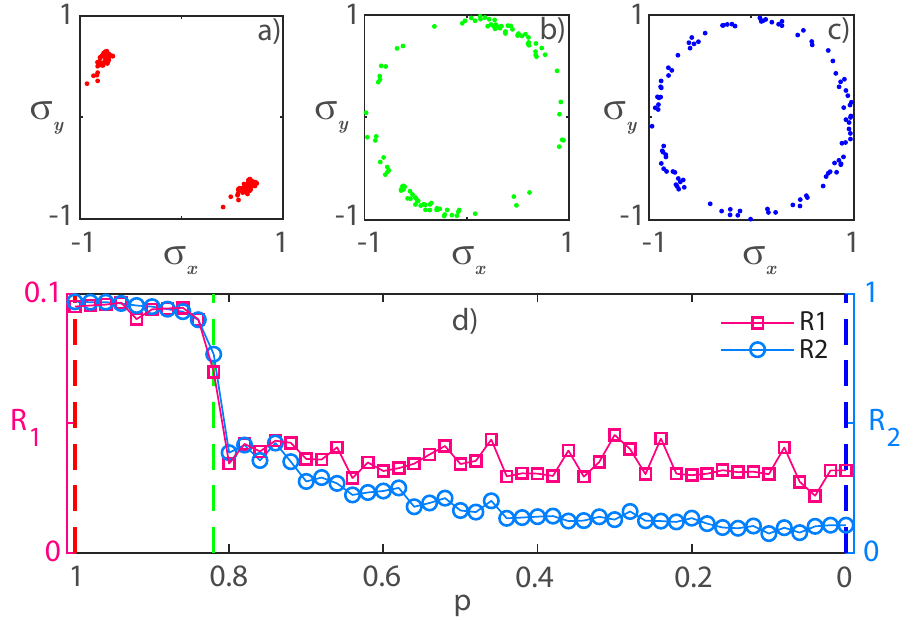}
	\caption{The resilience of the new state in sparser structures. Panel (d) reports $R_1$ (red curve with squares) and $R_2$ (blue curve with circles) vs. the fraction $p$ of remaining triangles (see text for definitions). Panel (a-c) show snapshots of the corresponding states of the oscillators in the plane ($\sigma_x = \cos\theta$, $\sigma_y=\sin\theta$) for three values of $p$ ($p=1, 0.84, 0$) marked by vertical dashed lines in panel (d) and colored with the same colors. Same stipulations for the phase representation as in the caption of Fig. 3.}
	\label{fig:04C}
\end{figure}

In summary, we have reported, both analytically and numerically, on the emergence of synchronization even when $\lambda_1$ and $\lambda_2$ are both negative. In other words, group interactions may lead to the emergence of synchronization, or in social terms to the emergence of agreement, even if all individuals in the network are contrarians, i.e., opposing the mainstream at all times. Microscopically, we have shown that this is due to imperfect symmetry breaking that splits the population into two groups with different phases, but with frequency synchrony in one.

Our research should find applicability in better understanding decision-making in human groups, especially where the decision space is multidimensional, such as in evolutionary settings involving the provisioning of public goods~\cite{perc_pr17} or the emergence of moral behavior~\cite{capraro_jrsi21}. Swarming in three dimensions~\cite{vicsek2012collective}, in particular where making swift decisions plays a key role even if not involving humans, such as in fish schools or murmurations under predation \cite{katz2011inferring, couzin2018synchronization}, might also benefit from the insights reported in our research.

This work was supported by the Slovenian Research Agency (Grant Nos. P1-0403 and J1-2457), by the Russian Federation Government (Grant No. 075-15-2019-1926), and by the program ``Leading Scientific Schools ”, Grant No. NSh-2540.2020.1.


%


\widetext
\clearpage

	\def\be{\begin{equation}}
	\def\ee{\end{equation}}

\def\bc{\begin{center}}
	\def\ec{\end{center}}
\def\bea{\begin{eqnarray}}
	\def\eea{\end{eqnarray}}

\def\ie{\textit{i.e.}}
\def\etal{\textit{et al.}}
\def\m{\vec{m}}
\def\G{\mathcal{G}}

\begin{center}
	\textbf{\large Supplementary Information: Contrarians synchronize beyond the limit of pairwise interactions}
\end{center}

\author{K. Kovalenko$^{+,*,}$}
\affiliation{Moscow Institute of Physics and Technology, Dolgoprudny, Moscow Region, 141701, Russian Federation}

\author{X. Dai$^{+,}$}
\affiliation{Northwestern Polytechnical University, Xi'an 710072, P. R. China}

\author{K. Alfaro-Bittner$^{+,}$}
\affiliation{Departamento de F{\'i}sica, Universidad T{\'e}cnica Federico Santa Mar{\'i}a, Av. Espa{\~n}a 1680, Casilla 110V, Valpara{\'i}so, Chile}

\author{A. M. Raigorodskii}
\affiliation{Moscow Institute of Physics and Technology, Dolgoprudny, Moscow Region, 141701, Russian Federation}
\affiliation{Adyghe State University, ul. Pervomaiskaya, 208, Maykop, 385000, Russia}
\affiliation{Moscow State University, Leninskie Gory, 1, Moscow, 119991, Russia}
\affiliation{Buryat State University, ul. Ranzhurova, 5, Ulan-Ude, 670000, Russia}

\author{M. Perc}
\affiliation{Faculty of Natural Sciences and Mathematics, University of Maribor, Koro{\v s}ka cesta 160, 2000 Maribor, Slovenia}
\affiliation{Department of Medical Research, China Medical University Hospital, China Medical University, Taichung 404332, Taiwan}
\affiliation{Complexity Science Hub Vienna, Josefst{\"a}dterstra{\ss}e 39, 1080 Vienna, Austria}
\affiliation{Alma Mater Europaea, Slovenska ulica 17, 2000 Maribor, Slovenia}

\author{S. Boccaletti}
\affiliation{Moscow Institute of Physics and Technology, Dolgoprudny, Moscow Region, 141701, Russian Federation}
\affiliation{Universidad Rey Juan Carlos, Calle Tulip\'an s/n, 28933 M\'ostoles, Madrid, Spain}
\affiliation{CNR - Institute of Complex Systems, Via Madonna del Piano 10, I-50019 Sesto Fiorentino, Italy}

\maketitle

In the following we provide an exact analytical treatment of the model, which allows one to obtain a self-consistent equation $r(\tilde{R}, \lambda_1, \lambda_2)$,and we will make use of part of the results which we have already described, for the pure 1-simplex and the pure 2-simplex interaction cases, in Ref.~\cite{csf2021}.

Let us remind that the behaviour of each unit $i$ in the system is described by the following differential equation

\bea
\dot{\bm\sigma_i} = \bm W_i \bm\sigma_i + \lambda_1[\bm\rho-(\bm\rho\cdot\bm\sigma_i)\bm\sigma_i] + \lambda_2(\bm\rho\cdot\bm\sigma_i)[\bm\rho-(\bm\rho\cdot\bm\sigma_i)\bm\sigma_i],
\label{eq:n-simplex_const}
\eea
where $\bm\rho=\frac{1}{N}\sum_{i=1}^{N}\bm\sigma_i$ is the mean vector of all oscillators.

As one is searching for stationary values of the synchronization parameter $R=|\bm\rho|$, the idea is to do it ith the following steps:
i) one replaces $\bm\rho$ with a fixed vector $\bm{\tilde{\rho}}$;
(ii)  after that, one computes the mean of the time-averaged projection of each unit onto $\bm{\tilde{\rho}}$, which can be denoted
as $r(\tilde{R},\lambda_1,\lambda_2)$, and (iii) then one solves the self-consistence equation
\bea
\tilde{R}=r(\tilde{R},\lambda_1,\lambda_2).
\label{eq:self-R}
\eea

\section{Stationary points for a single agent}

Since the matrices $\bm{W}_i$ are antisymmetric, for every node $i$ one can find an orthonormal basis in which $\bm{W}_i$ takes the block-diagonal form
\begin{equation}
	\label{eq:mat_even}
	{\bm W}_i=
	\begin{pmatrix}
		{\bm W}_i^{(1)} & 0 & \cdots & 0\\
		0 & {\bm W}_i^{(2)} & \cdots & 0\\
		\vdots & \vdots & \ddots & \vdots\\
		0 & 0 & \cdots & {\bm W}_i^{(D/2)}
	\end{pmatrix}
\end{equation}
if $D$ is even, and
\begin{equation}
	\label{eq:mat_odd}
	{\bm W}_i=
	\begin{pmatrix}
		{\bm W}_i^{(1)} & 0 & \cdots & 0&0\\
		0 & {\bm W}_i^{(2)} & \cdots & 0&0\\
		\vdots & \vdots & \ddots & \vdots&\vdots\\
		0 & 0&\cdots & {\bm W}_i^{((D-1)/2)}&0\\
		0 & 0&\cdots & 0&0\\
	\end{pmatrix}
\end{equation}
if $D$ is odd, where ${\bm W}_i^{(k)}$ is the $2\times 2$ matrix
\begin{equation*}
	{\bm W}_i^{(k)}=
	\begin{pmatrix}
		0 & \omega_i^{k}\\
		-\omega_i^{k} & 0
	\end{pmatrix}
\end{equation*}
and $\pm \mathbbm{i}\omega_i^{1},\cdots, \pm \mathbbm{i}\omega_i^{\lfloor D/2\rfloor}$ are the nonzero eigenvalues of $\bm{W}_i$.

In the same basis, one can represent $\bm{\sigma}_i$ and $\tilde{\bm \rho}$ as block vectors,
\bea
\label{eq:mat_even_2}
\bm\sigma_i=
\begin{pmatrix}
	\bm\sigma_i^{(1)} \\
	\bm{\sigma}_i^{(2)} \\
	\vdots\\
	\bm\sigma_i^{(D/2)}\\
\end{pmatrix},& \tilde{\bm\rho}=
\begin{pmatrix}
	\tilde{\bm\rho}^{(1)} \\
	\tilde{\bm{\rho}}^{(2)} \\
	\vdots\\
	\tilde{\bm\rho}^{(D/2)}\\
\end{pmatrix}
\eea
for $D$ even, and

\bea
\label{eq:mat_odd_2}
\bm\sigma_i=
\begin{pmatrix}
	\bm\sigma_i^{(1)} \\
	\bm{\sigma}_i^{(2)} \\
	\vdots\\
	{\bm \sigma}_i^{((D-1)/2)}\\
	\sigma_i^D\\
\end{pmatrix}, &
\tilde{\bm\rho}=
\begin{pmatrix}
	\tilde{\bm\rho}^{(1)} \\
	\tilde{\bm{\rho}}^{(2)} \\
	\vdots\\
	\tilde{\bm\rho}^{((D-1)/2)}\\
	\tilde{\rho}^{D}\\
\end{pmatrix}
\eea
for $D$ odd where we have denoted by $\bm\sigma_i^{(k)}$ and $\tilde{\bm\rho}^{(k)}$ the vectors
\bea
\bm\sigma_i^{(k)}=
\begin{pmatrix}
	\sigma_i^{2k-1}\\
	\sigma_i^{2k}
\end{pmatrix}, &
\tilde{\bm\rho}^{(k)}=
\begin{pmatrix}
	\tilde{\rho}^{2k-1}\\
	\tilde{\rho}^{2k}
\end{pmatrix}.
\eea

Recall that $\tilde{R}=|\tilde{\bm\rho}|$ and $\hat{\bm\rho}=\tilde{\bm \rho}/\tilde{R}$. Denote also $\hat{\bm\rho}^{(k)}=\tilde{\bm \rho}^{(k)}/\tilde{R}$. For each oscillator $i$, if one sets $\dot{\bm{\sigma}_i}=0$, Eq.~(\ref{eq:n-simplex_const}) transforms (in the corresponding basis) into the system of equations
\bea
\bm{W}_i^{(k)}\bm\sigma_i^{(k)}+
\left(\lambda_1 \tilde{R} + \lambda_2 \tilde{R}^2(\hat{\bm\rho}\cdot\bm\sigma_i)\right)
[\hat{\bm{\rho}}^{(k)}-(\hat{\bm\rho}\cdot\bm\sigma_i)\bm\sigma_i^{(k)}]={\bf 0}\nonumber
\eea
for all $k=1,\cdots,\frac{D}{2}$,

For odd $D$, one also has

\bea
\left(\lambda_1 \tilde{R} + \lambda_2 \tilde{R}^2(\hat{\bm\rho}\cdot\bm\sigma_i)\right)[\hat{\rho}^{D}-(\hat{\bm\rho}\cdot\bm\sigma_i)\bm\sigma_i^{D}]=0.\nonumber
\eea

By regrouping terms in these two equations, one gets

\bea
\bm{W}_i^{(k)}\bm\sigma_i^{(k)}-
\left(\lambda_1 \tilde{R}(\hat{\bm\rho}\cdot\bm\sigma_i) + \lambda_2 \tilde{R}^2(\hat{\bm\rho}\cdot\bm\sigma_i)^2\right)\bm\sigma_i^{(k)}=
\left(\lambda_1 \tilde{R} + \lambda_2 \tilde{R}^2(\hat{\bm\rho}\cdot\bm\sigma_i)\right)\hat{\bm{\rho}}^{(k)}
\label{uno}
\eea
and
\bea
\left(\lambda_1 \tilde{R}(\hat{\bm\rho}\cdot\bm\sigma_i) + \lambda_2 \tilde{R}^2(\hat{\bm\rho}\cdot\bm\sigma_i)^2\right)\bm\sigma_i^{D}=
\left(\lambda_1 \tilde{R}+ \lambda_2 \tilde{R}^2(\hat{\bm\rho}\cdot\bm\sigma_i)\right){\hat{\rho}}^{D}.\label{second}
\eea
Note that
\bea
\bm{W}_i^{(k)}\bm\sigma_i^{(k)}=\omega_i^k\begin{pmatrix} \sigma_i^{2k} \\-\sigma_i^{2k-1}\\\end{pmatrix},
\label{due}
\eea
hence
\bea
\left[\bm{W}_i^{(k)}\bm\sigma_i^{(k)}\right]\cdot  {\bm\sigma}_i^{(k)}&=&{\bf 0}
\label{trentaydos}
\eea
holds.
By taking the square of Eq.~(\ref{uno}) and by applying Eq.~(\ref{trentaydos}) to it, one obtains
{
	\bea
	\label{eq:a2+b2=c2-2-even}
	\left(\lambda_1 \tilde{R}(\hat{\bm\rho}\cdot\bm\sigma_i) + \lambda_2 \tilde{R}^2(\hat{\bm\rho}\cdot\bm\sigma_i)^2\right)^2
	\left| \bm\sigma_i^{(k)}\right|^2+
	\left| \bm{W}_i^{(k)}\bm\sigma_i^{(k)}
	\right|^2
	=
	\left(\lambda_1 \tilde{R} + \lambda_2 \tilde{R}^2(\hat{\bm\rho}\cdot\bm\sigma_i)\right)^2
	\left|{\hat{\bm\rho}}_i^{(k)}
	\right|^2.
	\eea}
One can further define $r_k$ and $l_k$ as follows:
\bea
r_k&=&\left|{\hat{\bm\rho}}_i^{(k)}
\right|=\sqrt{(\hat{\rho}_{2k-1})^2+(\hat{\rho}_{2k})^2},\nonumber \\
l_k&=&\left| \bm\sigma_i^{(k)}\right|=\sqrt{(\sigma_i^{2k-1})^2+(\sigma_i^{2k})^2}.
\label{eq:rk_lk}
\eea
Then, using Eq.~(\ref{due}) and Eq.~(\ref{eq:a2+b2=c2-2-even}), one can express the square of $l_k$ in the following way:
\begin{equation}
	{
		(l_k)^2
		=
		\frac{(r_k)^2\left(\lambda_1 \tilde{R} + \lambda_2 \tilde{R}^2(\hat{\bm\rho}\cdot\bm\sigma_i)\right)^2}{\left(\lambda_1 \tilde{R}(\hat{\bm\rho}\cdot\bm\sigma_i) + \lambda_2 \tilde{R}^2(\hat{\bm\rho}\cdot\bm\sigma_i)^2\right)^2+\left(\omega_k^{(i)}\right)^2}.}
	\label{eq:lk}
\end{equation}
Analogously, using Eq.~(\ref{second}) instead of Eq.~(\ref{uno}), one gets
\begin{equation}
	(l_D)^2=|\sigma_i^D|=\frac{\hat{\rho}_D^2}{(\hat{\bm\rho}\cdot\bm \sigma_i)^2}
	\label{eq:lD}
\end{equation}
for odd values of $D$ and $\hat{\bm\rho}\cdot\bm \sigma_i\neq 0$.
In the following $\lfloor x\rfloor$ indicates the floor function and $\delta_{x,y}$ the Kronecker delta.

Let us remind that $\bm\sigma_i$ lies on a unit sphere, so
{\bea
	|\bm\sigma_i|^2=\sum_{k=1}^{\lfloor D/2\rfloor}l_k^2+l_D^2\delta_{D,2\lfloor D/2\rfloor +1} =1
	\label{sigma2=1}
	\eea}
holds.

Using the expression for $l_k^2$ above, Eq.~(\ref{sigma2=1}) transforms into the following condition for the stationary point $\bm\sigma_i^F$:
{\bea
	1&=&\sum_{k=1}^{\lfloor D/2\rfloor} \frac{(\hat{\rho}_{2k-1}^2+\hat{\rho}_{2k}^2)\left(\lambda_1 \tilde{R} + \lambda_2 \tilde{R}^2(\hat{\bm\rho}\cdot\bm\sigma_i)\right)^2}{\left(\lambda_1 \tilde{R}(\hat{\bm\rho}\cdot\bm\sigma_i) + \lambda_2 \tilde{R}^2(\hat{\bm\rho}\cdot\bm\sigma_i)^2\right)^2+\left(\omega_i^{k}\right)^2}
	+\delta_{D,2\lfloor D/2\rfloor+1}\frac{\hat{\rho}_D^2}{(\hat{\bm\rho}\cdot\bm\sigma_i^F)^2}.
	\label{eq:n_self}
	\eea}

\section{Description of stable and unstable points}

Consider now a small perturbation $\dsi=\sigma_i-\sigma_i^F$ of $\sigma_i$ about $\sigma_i^F$, where $\sigma_i^F$ is a stationary point of $\sigma_i$. Then from Eq.~(\ref{eq:n-simplex_const}) we get
$$
\frac{d}{dt}\dsi = W_i (\bm\sigma_i^F+\dsi) + \lambda_1 (\tilde{\bm\rho}-(\tilde{\bm\rho}\cdot(\bm\sigma_i^F+\dsi))(\bm\sigma_i^F+\dsi))+
$$
$$
+\lambda_2 \left(\tilde{\bm\rho}\cdot(\bm\sigma_i^F+\dsi)\right)(\tilde{\bm\rho}-(\tilde{\bm\rho}\cdot(\bm\sigma_i^F+\dsi))(\bm\sigma_i^F+\dsi)).
$$
The point $\bm\sigma_i^F$ is stationary, so
$$
W_i \bm\sigma_i^F + \lambda_1 (\tilde{\bm\rho}-(\tilde{\bm\rho}\cdot\bm\sigma_i^F)\bm\sigma_i^F) + \lambda_2 (\tilde{\bm\rho}\cdot\bm\sigma_i^F)(\tilde{\bm\rho}-(\tilde{\bm\rho}\cdot\bm\sigma_i^F)\bm\sigma_i^F)=0,
$$
and it follows that
$$
\frac{d}{dt}\dsi = W_i \dsi - \lambda_1 \left((\tilde{\bm\rho}\cdot\bm\sigma_i^F)\dsi + (\tilde{\bm\rho}\cdot\dsi)\bm\sigma_i^F\right) -
$$
$$
-\lambda_2 (\tilde{\bm\rho}\cdot\bm\sigma_i^F)\left((\tilde{\bm\rho}\cdot\bm\sigma_i^F)\dsi + (\tilde{\bm\rho}\cdot\dsi)\bm\sigma_i^F\right)
+
$$
$$
+\lambda_2 (\tilde{\bm\rho}\cdot\dsi)(\tilde{\bm\rho}-(\tilde{\bm\rho}\cdot\bm\sigma_i^F)\bm\sigma_i^F) + O(\left\lVert\dsi\right\rVert_2^2).
$$
So
\bea
\frac{1}{2}\frac{d(\dsi\cdot\dsi)}{dt} =
\left(\dsi \cdot \frac{d}{dt}\dsi\right)
=
- \lambda_1 \left((\tilde{\bm\rho}\cdot\bm\sigma_i^F)(\dsi\cdot\dsi) + (\tilde{\bm\rho}\cdot\dsi)(\sigma_i^F\cdot\dsi)\right) -
\nonumber\\
\lambda_2 (\tilde{\bm\rho}\cdot\bm\sigma_i^F)((\tilde{\bm\rho}\cdot\bm\sigma_i^F)(\dsi\cdot\dsi)
+(\tilde{\bm\rho}\cdot\dsi)(\bm\sigma_i^F\cdot\dsi))+\\
\lambda_2 (\tilde{\bm\rho}\cdot\dsi)((\tilde{\bm\rho}\cdot\dsi)-(\tilde{\bm\rho}\cdot\bm\sigma_i^F)(\bm\sigma_i^F\cdot\dsi)) + o(\left\lVert\dsi\right\rVert_2^2).\nonumber
\eea
Remember that the vector $\bm\sigma_i$ always lies on the unit sphere, so the perturbation $\dsi$  is almost perpendicular to $\bm\sigma_i$, i.e. $(\bm\sigma_i^F \cdot \dsi)=o(\left\lVert\dsi\right\rVert_2)$. So, it follows that

\bea
\frac{1}{2}\frac{d(\dsi\cdot\dsi)}{dt} =
- (\dsi\cdot\dsi)\left[\lambda_1 (\tilde{\bm\rho}\cdot\bm\sigma_i^F) + \lambda_2\left((\tilde{\bm\rho}\cdot\bm\sigma_i^F)^2 - \left(\tilde{\bm\rho}\cdot\frac{\dsi}{\left\lVert\dsi\right\rVert_2}\right)^2\right)\right]
+o(\left\lVert\dsi\right\rVert_2^2).
\eea

From this latter equation one get the following criterion  - the stationary point is stable if and only if
\bea
\label{conditio}
\left[\lambda_1 (\tilde{\bm\rho}\cdot\bm\sigma_i^F) + \lambda_2\left((\tilde{\bm\rho}\cdot\bm\sigma_i^F)^2 - \left(\tilde{\rho}\cdot\frac{\dsi}{\left\lVert\dsi\right\rVert_2}\right)^2\right)\right] >0
\eea
on average in time for a small perturbation $\dsi$.

One should notice that when $(\tilde{\bm\rho}\cdot\bm\sigma_i^F)$ is close to $1$ (i.e., in proximity of the synchronous solution) and $\lambda_1\approx -\lambda_2$, then the first and second term on the left side of the inequality (\ref{conditio}) are close in absolute value but have different signs. As a consequence, their sum is close to zero, and therefore the criteria loses it's precision. In these conditions, a careful inspection of stability would imply considering also higher order terms in the equation for the perturbation $\dsi$.
\\

In the 2-dimensional case we have
$$
(\tilde{\bm\rho}\cdot\bm\sigma_i^F)^2 + \left(\tilde{\rho}\cdot\frac{\dsi}{\left\lVert\dsi\right\rVert_2}\right)^2 = 1,
$$
so for $D=2$ the stationary point is stable if and only if
\bea
\lambda_1 (\tilde{\bm\rho}\cdot\bm\sigma_i^F) + \lambda_2\left(2(\tilde{\bm\rho}\cdot\bm\sigma_i^F)^2 - 1\right)>0.
\eea

In the case of an odd $D$  and small $\lambda_1$ and $\lambda_2$, one can use the same approach as in Ref.~\cite{csf2021}, and gets
\bea
\left\langle\left(\tilde{\bm\rho}\cdot\frac{\dsi}{\left|\dsi\right|}\right)^2\right\rangle
\approx
\frac{1}{2}(1-(\hat{\bm\rho}\cdot\bm\sigma_i^F)^2)g_{\text{sign}(\lambda_2)}(D),\nonumber
\eea
where

\bea
g_{\pm}(D)=\int_{[0,+\infty]^{(D-1)/2}} {d}{\bf y} A_{\pm}({\bf y})\prod_{k=1}^{(D-1)/2}\left(\frac{1}{2}e^{-\frac{y_k}{2}}\right)\nonumber
\eea
and

\bea
A_+({\bf y})&=&\left[\max_{1\leq k\leq(D-1)/2} y_k\right] \left[\sum_{k=1}^{(D-1)/2}y_k\right]^{-1},
\\
A_-({\bf y})&=&\left[\min_{1\leq k\leq(D-1)/2} y_k\right] \left[\sum_{k=1}^{(D-1)/2}y_k\right]^{-1}.
\eea

In this case, one can assume that the stationary point is stable if and only if
\bea
\lambda_1 (\tilde{\rho}\cdot\sigma_i^F) +
\lambda_2\left((\tilde{\rho}\cdot\sigma_i^F)^2 - \frac{1}{2}(1-(\hat{\bm\rho}\cdot\bm\sigma_i^F)^2)g_{\text{sign}(\lambda_2)}(D)\right)&>&0.
\eea

Also for any dimension
$$
\left(\tilde{\rho}\cdot\frac{\dsi}{\left\lVert\dsi\right\rVert_2}\right)^2 \leq 1 - (\tilde{\rho}\cdot\sigma_i^F)^2,
$$
so for $\lambda_2>0$ if
\bea
\left[\lambda_1 (\tilde{\rho}\cdot\sigma_i^F) + \lambda_2\left(2(\tilde{\rho}\cdot\sigma_i^F)^2 - 1\right)\right] > 0,
\eea
then the stationary point is stable. This criterion is weaker than the criterion for $D=2$ but holds for any $D$. In our theory prediction we use the criterion for $D=2$ for greater even dimensions as well.

\section{Self-consistent equation}

One can define $h_{+}(\bm \omega,\hat{\bm\rho},\tilde{r},\lambda_1, \lambda_2)$
in the following way:
\begin{enumerate}
	\item to be the largest positive solution $\hat{\bm\rho}\cdot \bm\sigma_i^F$ of Eq.~(\ref{eq:n_self}) if such solution exists and is stable by the corresponding criterion;
	\item if there is no such a positive \textbf{stable} solution, then $h_{+}$ is taken to be the smallest negative solution of Eq.~(\ref{eq:n_self}) if such solution exists and is stable by the corresponding criterion;
	\item if there are no stable solutions, then $h_{+}(\bm \omega,\hat{\bm\rho},\tilde{r},\lambda_1, \lambda_2)$ is taken to be $0$.
\end{enumerate}

Analogously, one can define  $h_{-}(\bm \omega,\hat{\bm\rho},\tilde{r},\lambda_1, \lambda_2)$:
\begin{enumerate}
	\item to be the smallest \textbf{negative} solution $\hat{\bm\rho}\cdot \bm\sigma_i^F$ of Eq.~(\ref{eq:n_self}) if such solution exists and is stable by the corresponding criterion;
	\item if there is no such a negative \textbf{stable} solution, then $h_{-}$ is taken to be the largest \textbf{positive} solution of Eq.~(\ref{eq:n_self}) if such exists and is stable by the corresponding criterion;
	\item if there are no stable solutions, then $h_{-}(\bm \omega,\hat{\bm\rho},\tilde{r},\lambda_1, \lambda_2)$ is also taken to be $0$.
\end{enumerate}

With the above notation, the expression for $r(\tilde{R},\lambda_1, \lambda_2)$ appearing in the self-consistent Eq.~(\ref{eq:self-R}) reads as
$$
r(\tilde{R},\lambda_1,\lambda_2)=\int \left[\mu h_+(\bm\omega,\hat{\bm\rho},\tilde{R},\lambda_1, \lambda_2)  + (1-\mu)h_-(\bm\omega,\hat{\bm\rho},\tilde{R},\lambda_1, \lambda_2)\right] G(\bm{\omega}){ U}({\hat{\bm \rho}})d\bm\omega d\hat{\bm \rho},
$$

Once one knows the function $r$, the self-consistent values of $R$ for each given $\lambda_1$ and $\lambda_2$ can be obtained by solving the equation

\bea
\tilde{R}=r(\tilde{R},\lambda_1, \lambda_2).
\label{selffinale}
\eea

It should be pointed out that this equation always has at least one solution $\tilde{R}=0$.

A very important point (which actually is at the basis of the observed new effect of synchronized contrarians) to be discussed is that, for $D=2$ and $\lambda_2=0$ (i.e., for the classical case of the Kuramoto model with only pairwise interactions), if $\lambda_1$ is negative, then Eq.~(\ref{conditio}) tells us that a stationary point is stable if and only if it lies on the hemisphere opposite to the mean vector $\rho$. That means that the only solution of the self-consistent equation~(\ref{selffinale}) in this case is zero. This explains fully why synchronization of contrarians is actually prevented if only pairwise interactions are present.
In presence of 3-body interaction, instead, the second term on the left side of Eq.~(\ref{conditio}) depends on $(\tilde{\bm\rho}\cdot\bm\sigma^F_i)^2$, so this equality condition could be satisfied even for $\lambda_1<0$, $\lambda_2<0$ and for $(\tilde{\bm\rho}\cdot\bm\sigma^F_i)>0$.

Even though one chooses the stable solutions for individual units in the preceding analysis, the corresponding solutions of the self-consistence equation may still be unstable due to collective effects. One is, in fact, interested in the stable solutions for $\tilde{R}$. One natural stability criterion is
$$
\frac{\partial(r(\tilde{R},\lambda_1, \lambda_2) - \tilde{R})}{\partial\tilde{R}}<0.
$$
One can also introduce the following heuristic criterion. Let $R_0$ be a solution. If there is $R>R_0$ such that $r(R,\lambda)$ is significantly negative (for the purposes of our simulations, we used the cutoff $r(R,\lambda)<-0.05$), then one assumes that $R_0$ is an unstable stationary point. Otherwise, one relies on the criterion described above. If one finds that there are no stable nonzero solutions, one then assumes that the system is not synchronized and $R=0$.

\newpage

\section{Bifurcation diagrams}

In Fig.2 of the main text, one can see that the backward transition is abrupt in the region for positive $\lambda_2$ and negative  $\lambda_1$. Such a transition has the character of Explosive Desynchronization, and we here present the corresponding bifurcation analysis.

As a first point, one should notice that the system may have either one or an infinite number of stable states depending on the values of $\lambda_1$ and $\lambda_2$.
As a consequence, the system may display multi-stability, so that the values of the order parameter in the forward and backward transitions may depend on the initial conditions as well, and one can have, in principle, not one but multiple bifurcation diagrams.

Therefore, we focused our calculations on the following:
\begin{itemize}
	\item	the lower and the upper reachable bounds for stable states of the system;
	\item	the forward and backward transition, originated from a specific setting of initial conditions, i.e., from the specific values of $\mu=1$ for the backward transition and $\mu=0.5$ for the forward transition.
\end{itemize}

\begin{figure}[h!]
	\centering
	\includegraphics[scale=0.6]{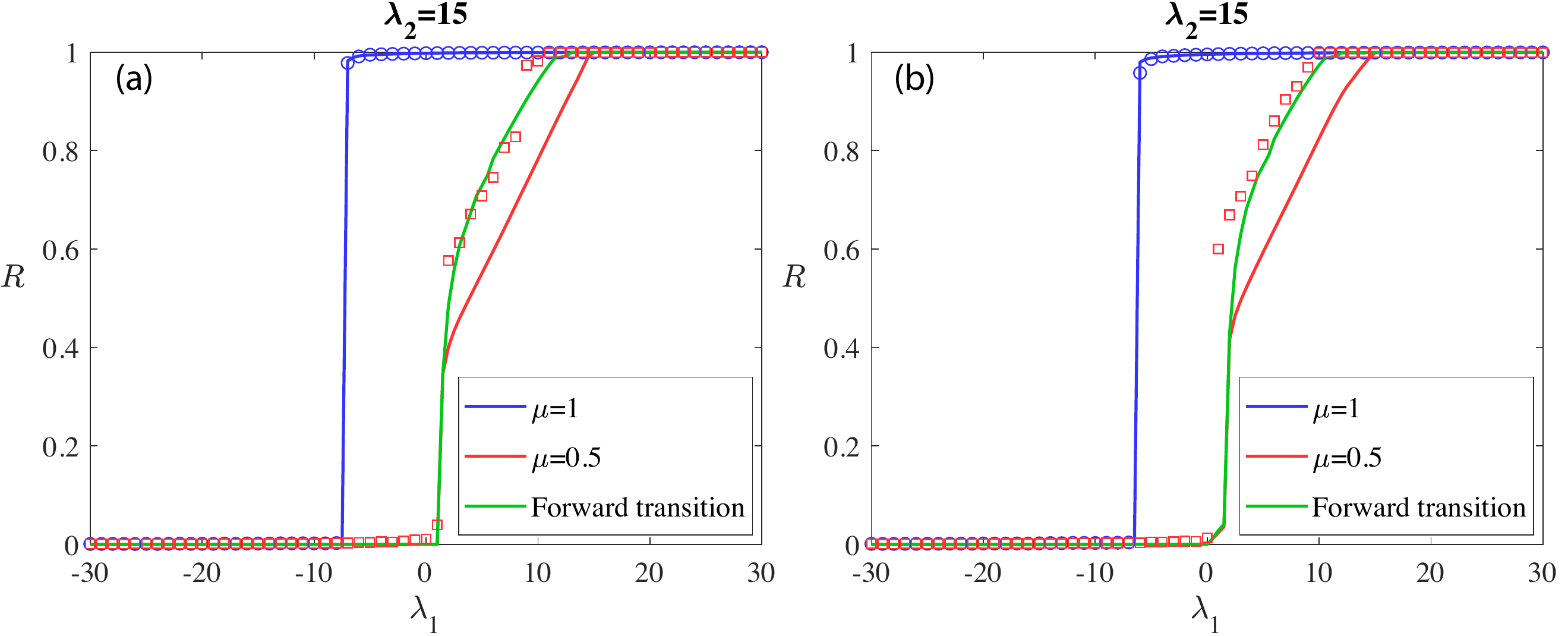}
	\caption{Bifurcation diagrams for $D=2$ (left panel) and $D=3$ (right panel), for fixed $\lambda_2=15$. In both cases, red solid lines correspond to theoretical lower reachable bound for stable states, blue solid lines correspond to theoretical upper reachable bounds for stable states, and green solid lines correspond to theoretical prediction for forward transition. Blue circles (red squares) refer to numerical simulations of the backward (forward) transition. Initial conditions (i.e., initial values of $\mu$) are specified in the insets.}
	\label{fig:my_label}
\end{figure}

We decided, instead, not to show the unstable solution branches since there may exist an infinite number of such solutions as well.
The results are reported in Figure~\ref{fig:my_label}.
Notice that the theoretical prediction for backward transition almost fully overlaps with the theoretical reachable upper bound, so we represent them with a unique solid line.

\end{document}